\documentclass[twocolumn,iop]{emulateapj}
\usepackage{apjfonts}
\usepackage{color}

\usepackage{amsmath,graphicx,longtable,hyperref}
\usepackage{natbib,threeparttable,rotating}
\usepackage{ulem}

\newcommand{\OI}{\makebox{[O\,{\sevenrm I}]\,}}

\newcommand{\etal}{et al.}
\newcommand{\hbeta}{H{$\beta$}}
\newcommand{\halpha}{H{$\alpha$}}

\newcommand{\CIV}{C{\sevenrm IV}}

\def\FeII{Fe\,{\sc ii}}
\def\MgII{Mg\,{\sc ii}}

\newcommand{\OII}{[O{\sevenrm\,II}]}

\def \OIII {[O\,{\sc iii}]}

\newcommand{\OIIIb}{[O{\sevenrm\,III}]\,$\lambda$5007}
\newcommand{\OIIIab}{[O{\sevenrm\,III}]\,$\lambda\lambda$4959,5007}

\newcommand{\SII}{[S{\sevenrm\,II}]}

   \font\sevenrm=cmr7 scaled 1000
\newcommand{\comments}[1]{}

\def\kms{{\rm km\,s^{-1}}}
\def\Rfe{R_{\rm FeII}}
\def\ergs{${\rm erg\,s^{-1}}$}

\begin{document}

\title{Rest-frame Optical Properties of Luminous $1.5<z<3.5$ Quasars: the \hbeta-\OIII\ region}

\author{Yue Shen$^{1}$} 

\altaffiltext{1}{Department of Astronomy and National Center for Supercomputing Applications, University of Illinois at Urbana-Champaign, Urbana, IL 61801, USA; shenyue@illinois.edu}

\shorttitle{Near-IR Spectroscopy of $1.5<z<3.5$ Quasars}
\shortauthors{Shen}

\begin{abstract}

We study the rest-frame optical properties of 74 luminous ($L_{\rm bol}=10^{46.2-48.2}\,{\rm erg\,s^{-1}}$), $1.5<z<3.5$ broad-line quasars with near-IR ($JHK$) slit spectroscopy. Systemic redshifts based on the peak of the \OIIIb\ line reveal that redshift estimates from the rest-frame UV broad emission lines (mostly \MgII) are intrinsically uncertain by $\sim 200\,{\rm km\,s^{-1}}$ (measurement errors accounted for). The overall full-width-at-half-maximum of the narrow \OIII\ line is $\sim 1000\,{\rm km\,s^{-1}}$ on average. A significant fraction of the total \OIII\ flux ($\sim 40\%$) is in a blueshifted wing component with a median velocity offset of $\sim 700\,\kms$, indicative of ionized outflows within a few kpc from the nucleus; we do not find evidence of significant \OIII\ flux beyond $\sim 10\,{\rm kpc}$ in our slit spectroscopy. The \OIII\ line is noticeably more asymmetric and weaker than that in typical less luminous low-$z$ quasars. However, when matched in quasar continuum luminosity, low-$z$ quasars have similar \OIII\ profiles and strengths as these high-$z$ systems. Therefore the exceptionally large width and blueshifted wing, and the relatively weak strength of \OIII\ in high-$z$ luminous quasars are mostly a luminosity effect rather than redshift evolution. The \hbeta-\OIII\ region of these high-$z$ quasars displays a similar spectral diversity and Eigenvector 1 correlations with anti-correlated \OIII\ and optical \FeII\ strengths, as seen in low-$z$ quasars; but the average broad \hbeta\ width is larger by 25\% than typical low-$z$ quasars, indicating more massive black holes in these high-$z$ systems. These results highlight the importance of understanding \OIII\ in the general context of quasar parameter space in order to understand quasar feedback in the form of \OIII\ outflows. The calibrated one-dimensional near-IR spectra are made publicly available, along with a composite spectrum.


\end{abstract}

\section{Introduction}\label{sec:intro}

Recent development of high-throughput near-IR spectrographs on moderate- to large-aperture telescopes has enabled the study of the rest-frame optical properties of high-redshift ($z>1.5$) quasars \citep[e.g.,][]{McIntosh_etal_1999,Yuan_Wills_2003,Shemmer_etal_2004,Netzer_etal_2004,Sulentic_etal_2004,Sulentic_etal_2006}. Over the past decade or so, the sample of high-redshift quasars with near-IR spectroscopy has grown considerably in size, and started to explore the statistical properties of the narrow-line regions (NLRs) and rest-frame optical broad-line regions (BLRs) of high-redshift quasars and their possible evolution from their low-redshift counterparts.

Near-IR spectroscopy of high-$z$ quasars provides a broad range of important applications, from estimating their black hole (BH) masses using the single-epoch virial BH mass estimators \citep[for a recent review, see][]{Shen_2013} based on the most reliable Balmer lines \citep[e.g.,][]{Dietrich_etal_2002,Dietrich_Hamann_2004,Netzer_etal_2004,Sulentic_etal_2004,Sulentic_etal_2006,Dietrich_etal_2009,Greene_etal_2010,Assef_etal_2011,Ho_etal_2012,Shen_Liu_2012,Runnoe_etal_2013b,Zuo_etal_2015,Brotherton_etal_2015,Plotkin_etal_2015,Shemmer_Lieber_2015,Saito_etal_2015}, to studying the sizes and kinematics of the NLR (usually utilizing the strong \OIIIab\ line) at $z>1.5$ \citep[e.g.,][]{Netzer_etal_2004,Nesvadba_etal_2008,Kim_etal_2013,Brusa_etal_2015,Perna_etal_2015,Carniani_etal_2015}. A wide wavelength coverage combining optical and near-IR spectroscopy of quasars is also valuable for constraining the spectral energy distribution (SED) of quasars and for testing accretion disk models \citep[e.g.,][]{Capellupo_etal_2015}.  

The sizes and kinematics of the NLR of low-$z$ Seyfert galaxies and quasars have been studied extensively in the past \citep[e.g.,][]{Mulchaey_etal_1996,Bennert_etal_2002,Schmitt_etal_2003,Bennert_etal_2006a,Bennert_etal_2006b}. These studies suggest that the NLR size increases with quasar luminosity, although there is evidence for an upper limit of $\sim 10\,{\rm kpc}$ on the NLR size for the most luminous quasars \citep[e.g.,][]{Netzer_etal_2004, Hainline_etal_2014}. In recent years, interests on NLR gas distributions and kinematics have been revived in the context of quasar-driven outflows and feedback from BH accretion. The \OIII\ emission in low-$z$ Seyferts and quasars often shows significant blueshifted velocity components indicative of outflows \citep[e.g.,][]{Heckman_etal_1981,Whittle_1985,Antonucci_2002}, and spatial extensions beyond $\sim {\rm kpc}$ scales, as inferred from spatially resolved slit or integral-field-unit (IFU) spectroscopy \citep[e.g.,][]{Stockton_MacKenty_1987,Crenshaw_Kraemer_2000,Nelson_etal_2000,Fu_Stockton_2009,Fischer_etal_2010,Villar-Martin_etal_2011,Greene_etal_2011,Shen_etal_2011b,Liu_etal_2013a,Liu_etal_2013b,Husemann_etal_2013}. Such kinematic studies of the NLR have been recently extended to $z>1.5$ for small samples with spatially-resolved near-IR spectroscopy \citep[e.g.,][]{Nesvadba_etal_2008,Harrison_etal_2012,Brusa_etal_2015,Perna_etal_2015,Carniani_etal_2015}, which suggest that quasar-driven outflows in ionized \OIII\ gas are common in luminous high-$z$ quasars. 

On the other hand, the rest-frame UV-to-optical regime of quasars displays a well-organized spectral diversity, which is ultimately connected to the fundamental properties of BH accretion. The most prominent feature of the quasar spectral diversity is a collection of quantities that all correlate with the strength of the optical \FeII\ emission known as Eigenvector 1 (EV1), discovered by \citet[][]{Boroson_Green_1992}. In the optical, one of the most important EV1 correlations is the anti-correlation between the strengths of \OIII\ and \FeII, even when quasar luminosity is fixed. EV1 has been the focus of quasar phenomenology for the past two decades \citep[e.g.,][]{Boroson_Green_1992,Wang_etal_1996,Boller_etal_1996,Brotherton_1996,Laor_1997,Laor_2000,Wills_etal_1999,Sulentic_etal_2000a,Marziani_etal_2001,Boroson_2002,Shang_etal_2003,Netzer_etal_2007,Sulentic_etal_2007,Dong_etal_2011,Shen_Ho_2014,Sun_Shen_2015}, as it provides important clues to understanding quasar accretion and feedback processes. It has been suggested that the main physical driver of EV1 is the Eddington ratio of the BH accretion \citep[e.g.,][]{Boroson_Green_1992,Sulentic_etal_2000a,Boroson_2002,Dong_etal_2011,Shen_Ho_2014,Sun_Shen_2015}, although other effects (such as orientation) may still play a minor role in affecting the observed strength of the \OIII\ lines \citep[e.g.,][]{Risaliti_etal_2011}. In addition to the EV1 correlation, the \OIII\ line also shows a Baldwin effect \citep{Baldwin_1977}, i.e., the restframe equivalent width (REW) of \OIII\ decreases as continuum luminosity increases \citep[e.g.,][]{Brotherton_1996,Zhang_etal_2011,Stern_Laor_2012a,Stern_Laor_2012b}, often accompanied by increasing flux in the blueshifted \OIII\ wing \citep[e.g.,][]{Zhang_etal_2013}. \citet{Shen_Ho_2014} presented a comprehensive analysis of the \OIII\ properties in low-$z$ quasars, and showed that the strength of the core \OIII\ component decreases with quasar luminosity and optical \FeII\ strength faster than the wing \OIII\ component, leading to overall broader and more blueshifted \OIII\ profiles as luminosity and \FeII\ strength increases. However, the blueshifted \OIII\ component appears to be a ubiquitous feature among quasars at different luminosities \citep[see figs.\ 2, E1 and E2 in][]{Shen_Ho_2014}. 

Building on these results on low-$z$ quasars, a natural step forward is to extend such studies to $z>1.5$ with near-IR spectroscopy to cover the rest-frame optical regime, and to investigate if these correlations involving \OIII\ and other optical lines exist at earlier times. However, compared with optical spectroscopy, near-IR spectroscopy of faint high-$z$ targets is much more expensive, and hence most of the earlier near-IR spectroscopic studies of high-$z$ quasars are still limited either by small sample statistics (of order ten objects or less) or by low spectral quality (i.e., low S/N, low spectral resolution, and/or limited spectral coverage). To enable a robust statistical study, high-quality near-IR spectroscopy for a large sample of high-$z$ quasars is therefore desirable. 

We have been conducting a near-IR spectroscopic survey of $1.5<z<3.5$ quasars to study their rest-frame optical properties. \citet{Shen_Liu_2012} presented a study on the correlations among virial mass estimators based on the UV broad lines and optical broad Balmer lines using 60 quasars at $z\sim 1.5-2.2$ from this survey. Here we present 14 additional quasars at $z\sim 3.5$ with new near-IR spectroscopy, and use a total of 74 quasars to study the rest-frame optical properties of these high-$z$ quasars, focusing on \OIII\ properties and EV1 correlations that involve the \hbeta-\OIII\ region. In \S\ref{sec:data} we describe the sample and spectral measurements, with a discussion on the systemic redshift estimation. We present our results in \S\ref{sec:results} and discussions in \S\ref{sec:disc}, and conclude in \S\ref{sec:con}. Throughout the paper we adopt a flat $\Lambda$CDM cosmology with $\Omega_0=0.3$ and $H_0=70\,{\rm km\,s^{-1}Mpc^{-1}}$. We use the REW to refer to the strength of a particular emission line. 

\section{Data and Spectral Measurements}\label{sec:data}

\subsection{Sample}

\begin{figure}
\centering
    \includegraphics[width=0.49\textwidth]{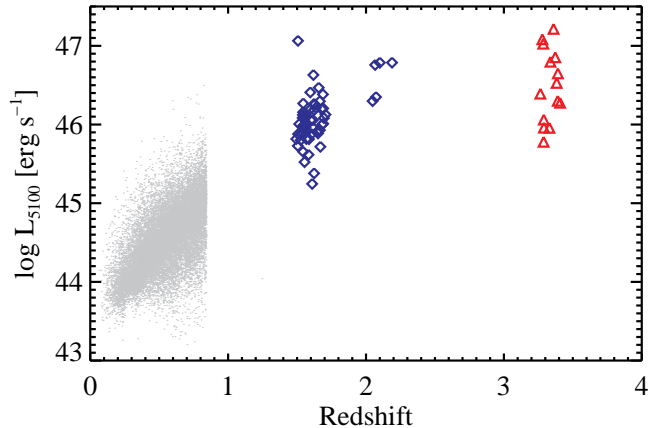}
    \caption{Distribution of our near-IR quasar sample (open symbols) in the redshift-luminosity plane. For comparison, we show the distribution of the low-$z$ SDSS sample from \citet{Shen_etal_2011} with \hbeta-\OIII\ coverage. }
    \label{fig:dist}
\end{figure}

\begin{table*}
\caption{New near-IR spectroscopic data}\label{table:basic}
\centering
\scalebox{1.0}{
\begin{tabular}{lcccccccccccc}
\hline\hline
Object Name &
RA (J2000) & DEC (J2000) & Plate & Fiber & MJD & $z_{\rm sys}$ & $i_{\rm PSF}$ & $J_{\rm 2MASS}$ & $H_{\rm 2MASS}$ & $K_{\rm s,2MASS}$ & NIR Obs. & Obs. UT\\
(1) & (2) & (3) & (4) & (5) & (6) & (7) & (8) & (9) & (10) & (11) & (12) & (13) \\
\hline
J0250$-$0757  & 02 50 21.77 & $-$07 57 50.0 & 0457 & 261 & 51901 & 3.3376 & 17.99 & 16.7  & 16.1  & 15.9  & FIRE & 131228 \\
J0259$+$0011 & 02 59 05.64 & $+$00 11 21.9 & 0411 & 398 & 51817 & 3.3724 & 17.75 & 16.2 (16.43) & 15.5 (15.83) & 15.2 (15.18) & FIRE & 131230 \\
J0304$-$0008 & 03 04 49.86 & $-$00 08 13.5 & 0411 & 153 & 51817 & 3.2859 & 17.47  & 16.3 (16.22) & 15.7 (15.86) & 15.3 (15.21) & FIRE & 131229 \\
J0352$-$0517 & 03 52 20.70 & $-$05 17 02.7 & 2071 & 353 & 53741 & 3.2892 & 18.73  & 17.6 &  0.0 &  0.0 & FIRE &  131230 \\
J0810$+$0936 & 08 10 11.97 & $+$09 36 48.3 & 2421 & 214 & 54153 & 3.3906 & 18.35 & 17.4 &  0.0 & 16.2 & FIRE & 131229 \\
J0843$+$0750 & 08 43 12.64 & $+$07 50 29.3 & 1298 & 376 & 52964 & 3.2648 & 19.22 & 16.9 (17.82) &  0.0 (17.24) &  0.0 (16.50) & FIRE & 131229 \\
J0844$+$0503 & 08 44 01.96 & $+$05 03 57.9 & 1188 & 464 & 52650 & 3.3603 & 17.09 & 15.4 (15.46) & 14.9 (15.00) & 14.2 (14.24) & FIRE & 131228 \\
J0910$+$0237 & 09 10 54.79 & $+$02 37 04.6 & 0566 & 014 & 52238 & 3.2902 & 18.56 & 17.3 (17.57) & 16.1 (17.13) &  0.0 (16.57) & FIRE & 131230 \\
J0942$+$0422 & 09 42 02.05 & $+$04 22 44.5 & 0570 & 427 & 52266 & 3.2790 & 17.18 & 15.9 (15.75) & 15.3 (15.23) & 14.6 (14.58) & FIRE & 131228 \\
J0953$+$0336 & 09 53 33.71 & $+$03 36 23.7 & 0571 & 114 & 52286 & 3.2881 & 19.85 & 17.4 (18.27) &  0.0 (17.74) &  0.0 (17.23) & FIRE & 131230 \\
J0954$+$0915 & 09 54 34.94 & $+$09 15 19.6 & 1306 & 052 & 52996 & 3.4076 & 18.67 & 17.4 (17.44) & 16.5 (17.11) &  0.0 (16.61) & FIRE &  131230 \\
J1019$+$0254 & 10 19 08.27 & $+$02 54 31.9 & 0503 & 456 & 51999 & 3.3829 & 18.03 & 16.7 (16.72) & 16.2 (16.38) & 15.5 (15.76) & FIRE & 131229 \\
J1034$+$0358 & 10 34 56.31 & $+$03 58 59.4 & 0576 & 026 & 52325 & 3.3918 & 17.69 & 16.4 (16.43) & 15.8 (15.95) & 15.4 (15.36) & FIRE & 131229 \\
J2238$-$0921 & 22 38 19.77 & $-$09 21 06.0 & 0722 & 190 & 52224 & 3.3300 & 17.81  & 16.7 & 16.2 & 15.5 & FIRE & 130529 \\
\hline
\hline\\
\end{tabular}
}
\begin{tablenotes}
      \small
      \item NOTE. --- Summary of the 14 new SDSS quasars at $z\sim 3.3$ for which we have
obtained near-infrared spectroscopy. Columns (4)-(6): plate, fiber and MJD
of the optical SDSS spectrum for each object; (7): systemic redshift
determined from the near-IR spectrum (see \S\ref{sec:spec_mea}); (8): SDSS $i$-band PSF magnitudes; (9)-(11):
2MASS magnitudes (Vega) and UKIDSS \citep{Lawrence_etal_2007} magnitudes (Vega) in the parentheses when available; (12): instrument for the near-IR spectroscopy; (13): UT dates of the near-IR observations. Note that here the 2MASS magnitudes
were taken from \citet{Schneider_etal_2010}, where aperture photometry was
performed upon 2MASS images to detect faint objects, hence these near
infrared data go beyond the 2MASS All-Sky and ``$6\times$'' point source
catalogs \citep[see][ for details]{Schneider_etal_2010}; zero values indicate non-detections.
\end{tablenotes}
\end{table*}

Our sample of quasars with near-IR spectroscopy consists of the 60 quasars at $z\sim 1.5-2.2$ from \citet{Shen_Liu_2012}, and 14 additional quasars at $z\sim 3.3$ for which we have obtained near-IR spectroscopy with the Folded-port InfraRed Echellett \citep[FIRE;][]{Simcoe_etal_2010} on the 6.5 m Magellan-Baade telescope during two runs in May and Dec, 2013. Table \ref{table:basic} summarizes the basic information of the 14 new $z\sim 3.3$ quasars. The data reduction and flux calibration of the new FIRE spectroscopy followed the same procedure as described in \citet{Shen_Liu_2012}. All 74 quasars have simultaneous $JHK$ coverage in the near-IR. These quasars were selected from the SDSS DR7 quasar catalog \citep{Shen_etal_2011} with good optical spectra covering the \CIV\ line and in redshift windows of $z\sim 1.5, 2.1, 3.3$ such that the \hbeta-\OIII\ region can be covered in the $JHK$ bands in the near-IR. Requiring them to have good quality (S/N per pixel$\gtrsim 10$) SDSS spectra preferentially selects high-luminosity quasars at these redshifts, but the resulting sample still covers a range of spectral diversities in the emission line properties. Most of these quasars are radio-quiet \citep[][]{Shen_etal_2011}. 

Fig.\ \ref{fig:dist} shows the distribution of our sample in the redshift-luminosity plane, compared to the low-$z$ quasar sample drawn from SDSS DR7 \citep{Shen_etal_2011} with optical spectroscopy covering the \hbeta-\OIII\ region. We have applied an average correction for host contamination in the rest-frame 5100\,\AA\ luminosities for the low-$z$ comparison sample, using the empirical formula in \citet[][eqn.\ 1]{Shen_etal_2011}. There is no need to apply this correction for the luminous quasars in our near-IR sample. 

The reduced and calibrated 1d near-IR spectra for all 74 quasars used in this study are available in ASCII format in the online version of the paper.   

\subsection{Spectral Measurements}\label{sec:spec_mea}

We use functional fits to measure the continuum and emission line properties of our near-IR quasar sample following earlier work \citep[e.g.,][]{Shen_etal_2008,Shen_etal_2011} in the \hbeta-\OIII\ region covered by near-IR spectroscopy. In short, we fit a local power-law continuum plus an empirical optical \FeII\ template \citep{Boroson_Green_1992} to the wavelength regions just outside the \hbeta-\OIII\ complex to form a pseudo-continuum. We then subtract this pseudo-continuum model from the original spectrum, and fit a number of Gaussian functions (in logarithmic wavelength space) to model the broad and narrow emission lines. We used 3 Gaussians to describe the broad \hbeta\ and 1 Gaussian to describe the narrow \hbeta. The \OIIIab\ doublet was each modeled by two Gaussians, one for the ``core'' component and one for the blueshifted ``wing'' component. During the fit, the velocity shift and dispersion of the narrow \hbeta\ component are tied to those of the ``core'' component of \OIII. 

\begin{figure}
\centering
    \includegraphics[width=0.49\textwidth]{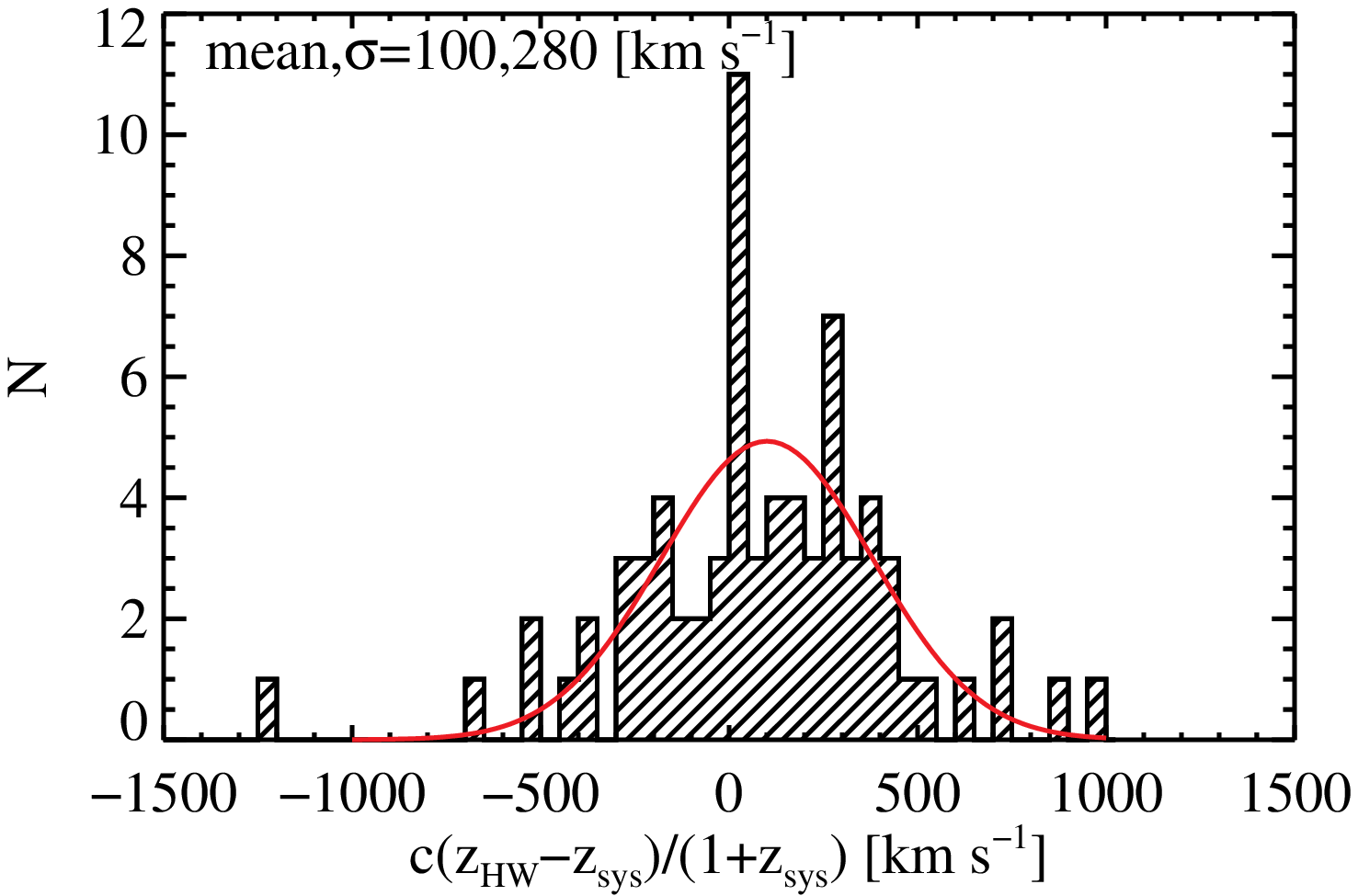}
    \caption{Histogram of the differences (with measurement errors) between the HW redshifts based on rest-frame UV lines (mostly \MgII) and the systemic redshifts based on \OIII\ (or \hbeta) for our near-IR sample. The red line is a best-fit Gaussian to the distribution, with the mean and dispersion shown. The HW redshifts are on average larger than the \OIII\ (\hbeta)-based systemic redshifts by $\sim 100\,{\rm km\,s^{-1}}$.}
    \label{fig:zdiff}
\end{figure}

We determine the systemic redshift $z_{\rm sys}$ using the model peak of the full \OIIIb\ profile; in cases where \OIII\ is not covered or if its S/N is poor, we use the model peak from the \hbeta\ line. Fig.\ \ref{fig:zdiff} compares these systemic redshifts with those reported by \citet{Hewett_Wild_2010} ($z_{\rm HW}$), which were based on cross-correlations of rest-frame UV lines (mostly \MgII\ for most of our objects) with quasar templates and empirical corrections for the velocity offsets between UV lines and the systemic redshifts. There is a small mean offset of $\sim 100\,{\rm km\,s^{-1}}$ between $z_{\rm HW}$ and $z_{\rm sys}$ and a dispersion of $\sim 280\,{\rm km\,s^{-1}}$ between the two redshifts. The reported measurement uncertainties in the Hewett \& Wild redshifts are typically $\sim 180\,{\rm km\,s^{-1}}$, while the typical measurement uncertainty in our systemic redshift estimates is $\sim 60\,{\rm km\,s^{-1}}$. Subtracting the typical measurement uncertainties, there is a residual difference of $\sim 200\,{\rm km\,s^{-1}}$ in the two sets of redshifts, which reflects the systematic uncertainty in estimating the systemic redshifts based on rest-frame UV lines (mostly \MgII\ for the bulk of our near-IR sample). This systematic uncertainty in \MgII-based redshifts is consistent with that inferred from comparing \MgII\ and \OIII-based redshifts in low-$z$ SDSS quasars using the spectral measurements in \citet[][]{Shen_etal_2011}. The refined systemic redshifts for the near-IR sample are important for deriving the composite spectrum and studying the average line profile of these objects.

We note that for the bulk of the population the peak of the full \OIII\ is consistent to within $\sim 50\,{\rm km\,s^{-1}}$ with the systemic redshifts based on stellar absorption lines \citep[e.g.,][]{Hewett_Wild_2010, Bae_Woo_2014} or low-ionization lines such as \SII\ \citep[e.g.,][]{Zhang_etal_2011} or \OII\ \citep[e.g.,][]{Shen_Ho_2014}, although rare individual objects could have a large blueshifted velocity offset in the \OIII\ peak \citep[e.g.,][]{Zamanov_etal_2002,Boroson_2005,Komossa_etal_2008}. However, if a single Gaussian were fit to the overall \OIII\ profile, the potential blueshifted ``wing'' component could bias the redshift estimation based on the Gaussian peak.  

On the other hand, \citet{Hewett_Wild_2010} suggested that, based on the analysis of low-$z$ quasar spectra, the \OIII\ centroid measured above the 50 per cent peak-height level\footnote{This is slightly different from our approach of measuring the \OIII\ peak velocity from model fits. }, is on average blueshifted from systemic (defined by Ca II K line at 3934.8\,\AA) by $\sim 45\,{\rm km\,s^{-1}}$. Taken this mean offset of \OIII\ into account, the comparison shown in Fig.\ \ref{fig:zdiff} suggests that the \MgII-based redshifts in \citet{Hewett_Wild_2010} only mildly overestimate the systemic redshifts by $\sim 50\,{\rm km\,s^{-1}}$ on average for our quasars.

\begin{figure}
\centering
    \includegraphics[width=0.48\textwidth]{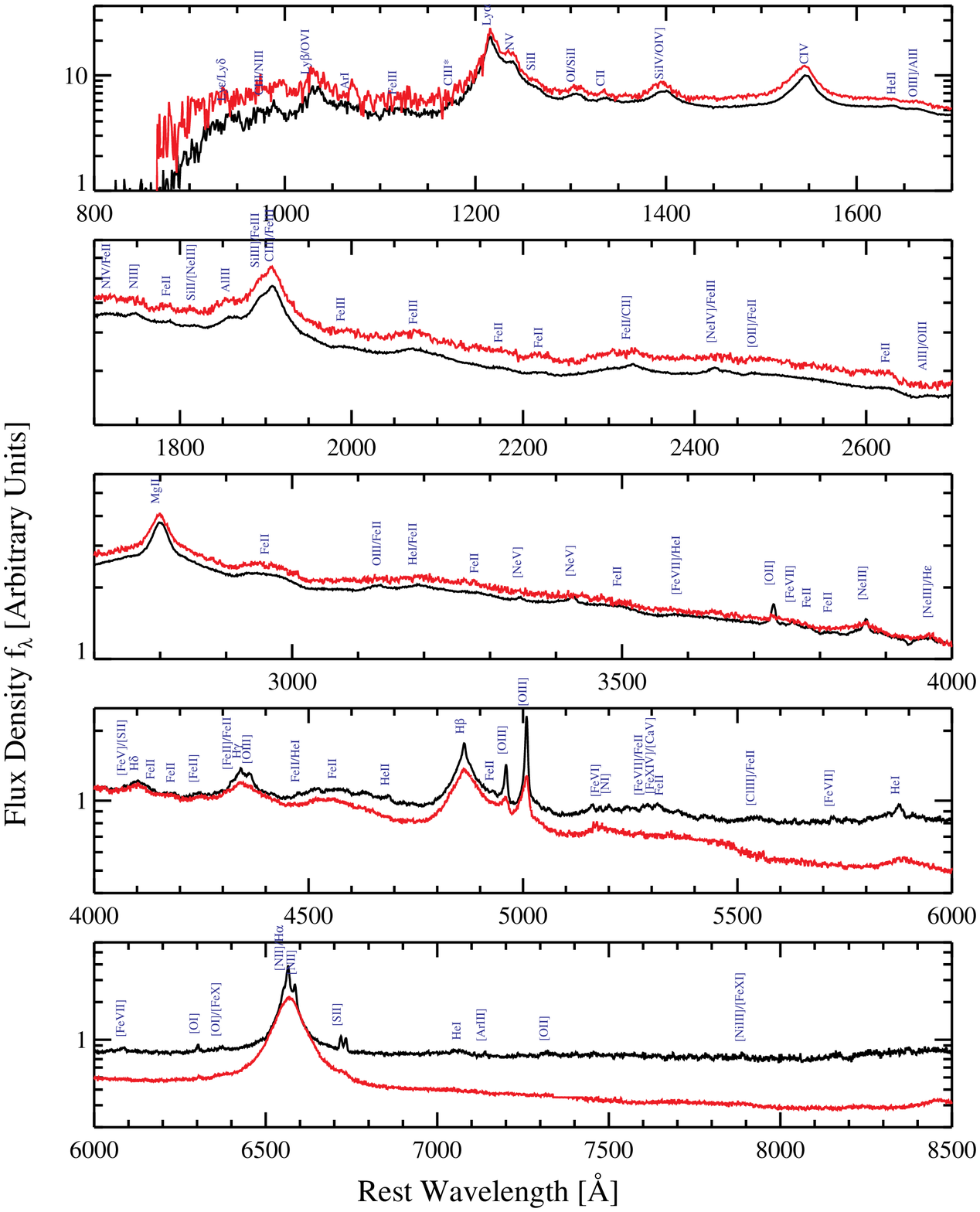}
    \caption{Median composite spectra for our near-IR sample (red) and for SDSS quasars from \citet[][black]{VandenBerk_etal_2001}. Other than the apparently broader emission line profiles and the bluer continuum (see text), the composite spectrum for the high-$z$ quasars is similar to that of general SDSS quasars. The full composite spectrum for the near-IR sample is tabulated in Table \ref{table:composite}. }
    \label{fig:fullspec}
\end{figure}

\begin{figure*}
\centering
    \includegraphics[width=0.47\textwidth]{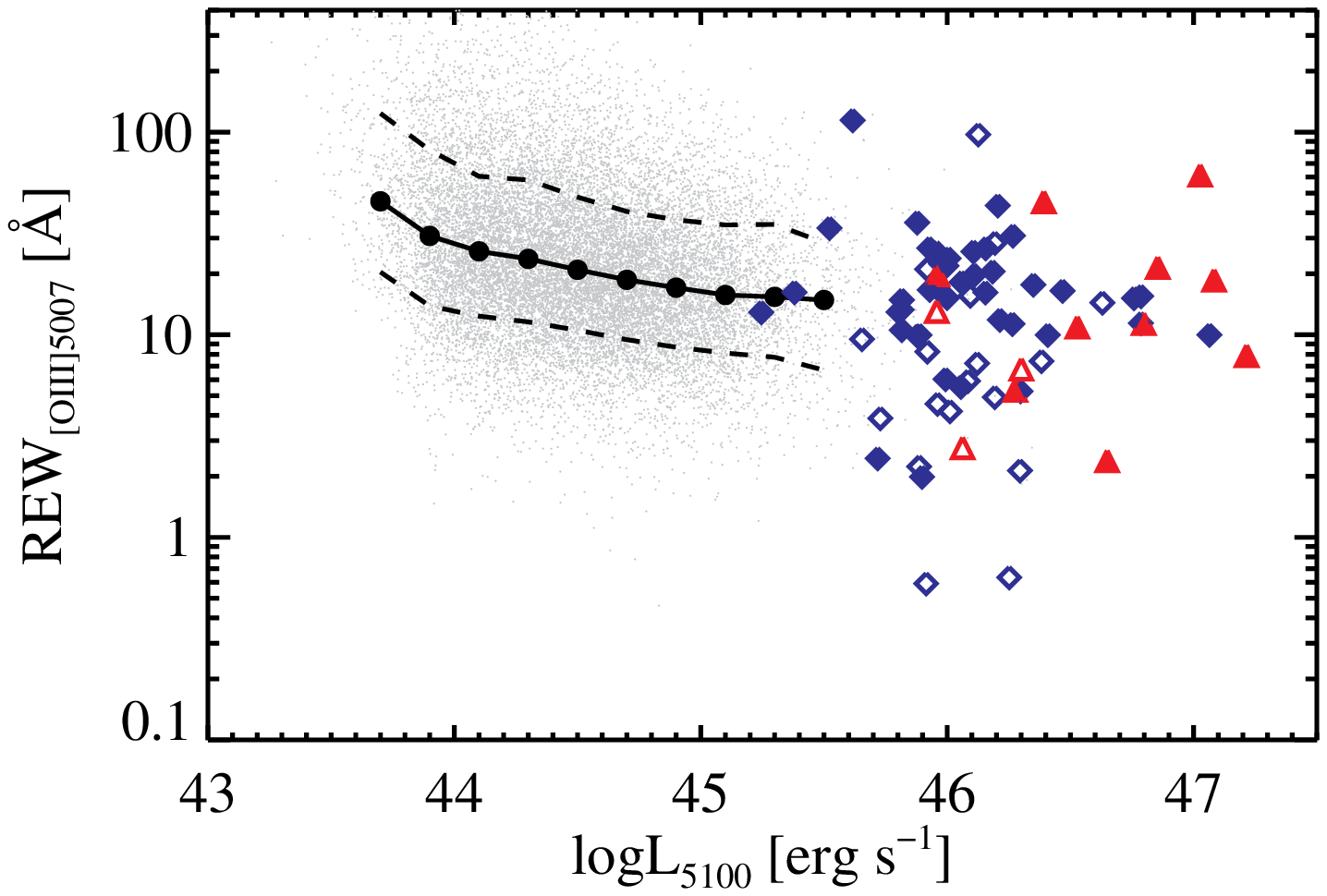}
    \includegraphics[width=0.47\textwidth]{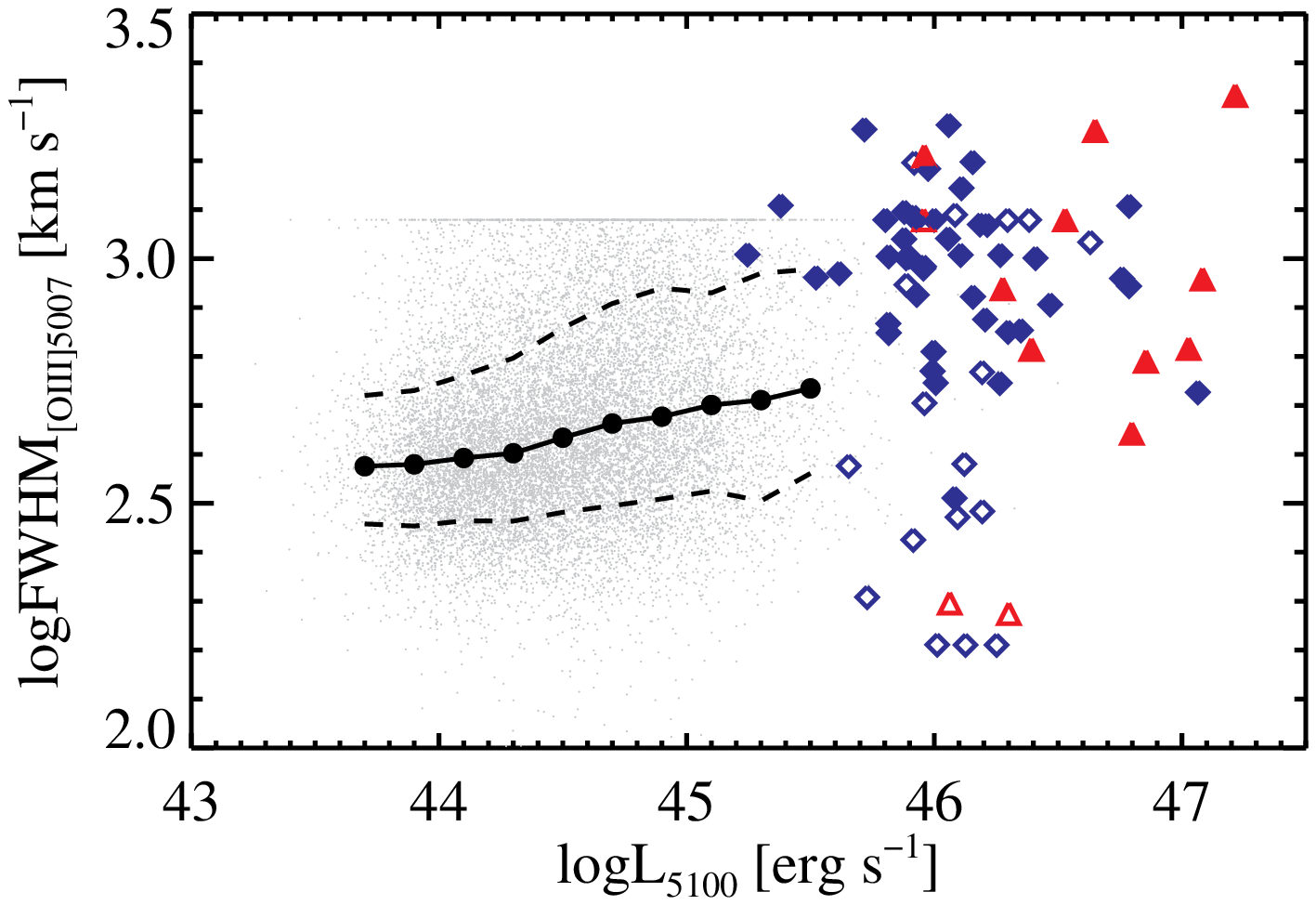}
    \caption{Comparisons between our high-$z$ quasars (diamonds and triangles) and the $z<1$ SDSS quasars (gray dots), where the latter measurements are from \citet{Shen_etal_2011}. {\it Left:} the total \OIII\ equivalent width as a function of the rest-frame 5100\,\AA\ continuum luminosity. {\it Right:} the FWHM of the entire \OIII\ profile as a function of $L_{5100}$. Symbol notations are the same as in Fig.\ \ref{fig:dist}. The lines are the 16th, 50th and 84th percentiles of the distribution of low-$z$ SDSS quasars. The open symbols are those with low-quality \OIII\ detections (S/N$<3$). We caution that individual measurements for the near-IR sample can be quite noisy (see Table \ref{table:fits}). The pileup of gray points at the top in the right panel is due to the upper limit of $1200\,{\rm km\,s^{-1}}$ imposed in the \OIII\ fits in \citet{Shen_etal_2011}.  }
    \label{fig:oiii}
\end{figure*}

We measure the continuum luminosity at rest-frame 5100\,\AA\ and the emission line properties (such as the rest-frame equivalent width REW and FWHM) using the model fits. To estimate measurement errors, we use a Monte Carlo approach \citep[e.g.,][]{Shen_etal_2008,Shen_etal_2011}: for each object we perturb the original spectrum by adding artificial noise using the reported spectral error array to generate a mock spectrum and perform the same fit on it; we repeat the process for 50 realizations of mock spectra and record the measurements for each realization; the nominal 1$\sigma$ measurement errors are then estimated as the semi-amplitude of the range enclosing the 16th and 84th percentiles of the distribution from the mock spectra. We provide a fits catalog of the spectral measurements in the \hbeta-\OIII\ region for our near-IR sample and documented the catalog in Table \ref{table:fits}. Additional properties of these quasars can be found in the \citet{Shen_etal_2011} catalog.

\begin{figure}
    \includegraphics[width=0.49\textwidth]{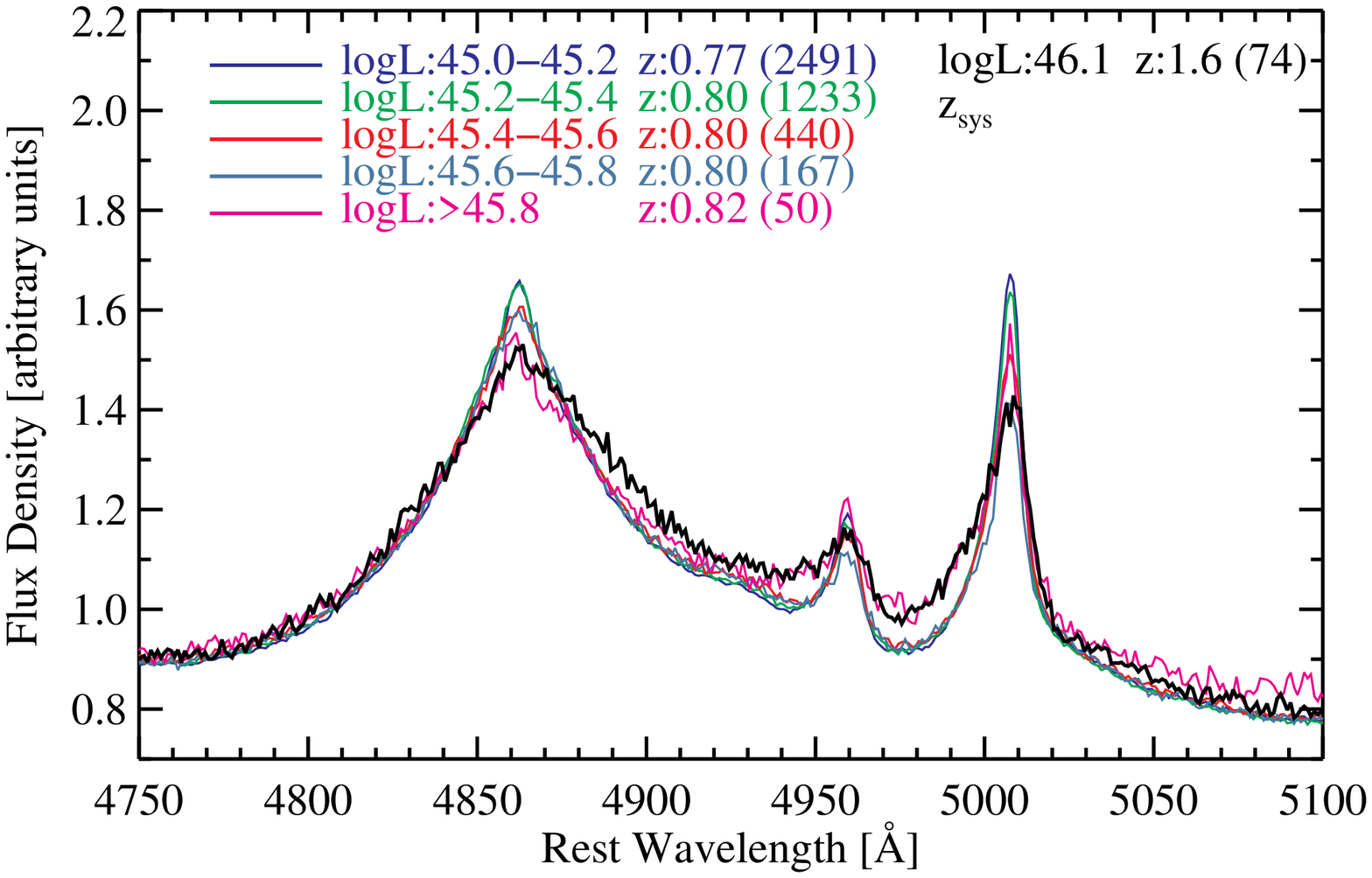}
    \includegraphics[width=0.49\textwidth]{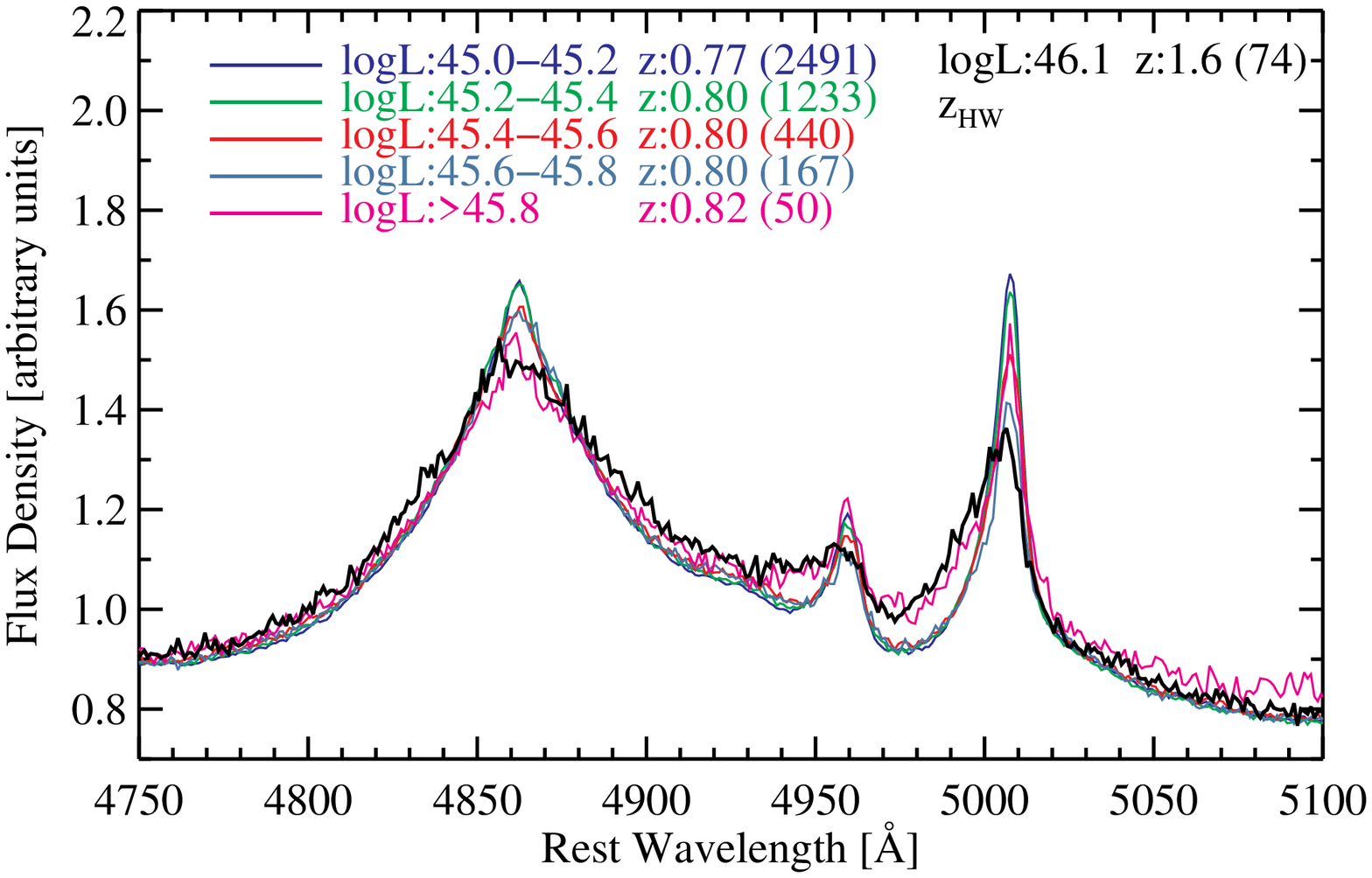}
    \caption{Median composite spectra of the $z>1.5$ near-IR quasar sample (black lines) and comparison with the composite spectra of the most luminous low-$z$ SDSS quasars in different luminosity bins. \textit{Top:} Using the estimated systemic redshifts $z_{\rm sys}$ for the near-IR sample. \textit{Bottom:} Using the redshifts reported by \citet{Hewett_Wild_2010} for the near-IR sample. The Hewett \& Wild redshifts appear to be overestimated on average for the near-IR sample, and have an additional scatter relative to the systemic redshifts from \OIII\ (and \hbeta). }
    \label{fig:coadd}
\end{figure}

\section{Results}\label{sec:results}

\begin{table}
\caption{Composite Spectrum}\label{table:composite}
\centering
\scalebox{1.0}{
\begin{tabular}{lccc}
\hline\hline
Rest Wavelength & Flux & Flux Error & $N_{\rm obj}$ \\
(1) & (2) & (3) & (4) \\
\hline
800.5 & 0.000  & 0.000 & 0 \\
801.5 & 0.000  & 0.000 & 0 \\
\hline
\hline\\
\end{tabular}
}
\begin{tablenotes}
      \small
      \item NOTE. --- Median composite spectrum for our near-IR quasar sample. Wavelengths are in units of \AA. Flux and flux error units are arbitrary. The last column indicates how many objects contributed to the median composite at each wavelength pixel. 
\end{tablenotes}
\end{table}

\begin{table*}
\caption{Spectral Measurements}\label{table:fits}
\centering
\scalebox{1.0}{
\begin{tabular}{llll}
\hline\hline
Column & Format & Units & Description \\
\hline
OBJNAME & A10 & -- & Object Name \\
PLATE & LONG & -- & Plate number of the SDSS spectrum \\
FIBER & LONG & -- & Fiber ID of the SDSS spectrum \\
MJD & LONG & -- &  Modified Julian Date of the SDSS spectrum \\
RA & DOUBLE & degree & J2000 Right ascension \\
DEC & DOUBLE & degree &  J2000 Declination \\
ZHW & FLOAT & -- & Redshfit from \citet{Hewett_Wild_2010} \\
ZHW\_ERR & FLOAT & -- & Measurement error in ZHW \\
ZSYS & FLOAT & -- & Systemic redshift \\
ZSYS\_ERR & FLOAT & -- & Measurement error in ZSYS \\
LOGL5100 &       DOUBLE     &      [${\rm erg\,s^{-1}}$] & Continuum luminosity at rest-frame 5100\,\AA \\
LOGL5100\_ERR &   DOUBLE   & [${\rm erg\,s^{-1}}$] &  Measurement error in LOGL5100  \\
FWHM\_BROAD\_HB  & DOUBLE   &     $\kms$ & FWHM of the broad \hbeta\ component \\
FWHM\_BROAD\_HB\_ERR & DOUBLE  &   $\kms$   & Measurement error in FWHM\_BROAD\_HB \\
EW\_BROAD\_HB  &   DOUBLE & \AA           & Rest-frame equivalent width of the broad \hbeta\ component \\
EW\_BROAD\_HB\_ERR & DOUBLE   & \AA    & Measurement error in EW\_BROAD\_HB \\
LOGL\_BROAD\_HB  & DOUBLE  &   [\ergs]     & Luminosity of the broad \hbeta\ component \\
LOGL\_BROAD\_HB\_ERR  & DOUBLE  & [\ergs] &  Measurement error in LOGL\_BROAD\_HB \\
LOGL\_NARROW\_HB  & DOUBLE  & [\ergs] & Luminosity of the narrow \hbeta\ component \\
LOGL\_NARROW\_HB\_ERR & DOUBLE  & [\ergs] & Measurement error in LOGL\_NARROW\_HB \\
EW\_OIII\_5007  &   DOUBLE  & \AA    & Rest-frame equivalent width of the entire \OIIIb\ line \\
EW\_OIII\_5007\_ERR & DOUBLE  & \AA  & Measurement error in EW\_OIII\_5007\_ERR \\
LOGL\_OIII\_5007  & DOUBLE & [\ergs]  & Luminosity of the entire \OIIIb\ line \\
LOGL\_OIII\_5007\_ERR & DOUBLE  & [\ergs] & Measurement error in LOGL\_OIII\_5007 \\
LOGL\_OIII\_5007C & DOUBLE  & [\ergs] & Luminosity of the core component of \OIIIb\ \\
LOGL\_OIII\_5007C\_ERR & DOUBLE  & [\ergs]  & Measurement error in LOGL\_OIII\_5007C \\
LOGL\_OIII\_5007W & DOUBLE  & [\ergs] & Luminosity of the wing component of \OIIIb\ \\
LOGL\_OIII\_5007W\_ERR & DOUBLE   & [\ergs]    & Measurement error in LOGL\_OIII\_5007W \\
FWHM\_OIII\_5007 & DOUBLE  & $\kms$  & FWHM of the entire \OIIIb\ line \\
FWHM\_OIII\_5007\_ERR & DOUBLE  & $\kms$  & Measurement error in FWHM\_OIII\_5007 \\
FWHM\_OIII\_5007C & DOUBLE  &  $\kms$ & FWHM of the core component of \OIIIb\ \\
FWHM\_OIII\_5007C\_ERR & DOUBLE  & $\kms$ & Measurement error in FWHM\_OIII\_5007C  \\
FWHM\_OIII\_5007W & DOUBLE & $\kms$ & FWHM of the wing component of \OIIIb\ \\
FWHM\_OIII\_5007W\_ERR & DOUBLE  & $\kms$  & Measurement error in FWHM\_OIII\_5007W \\
VOFF\_OIII\_5007W & DOUBLE & $\kms$ & Velocity offset of the wing component of \OIIIb\ from systemic (negative means blueshift) \\
VOFF\_OIII\_5007W\_ERR & DOUBLE & $\kms$  & Measurement error in VOFF\_OIII\_5007W \\
EW\_FE\_4434\_4684 & DOUBLE  & \AA & Rest-frame equivalent width of the optical \FeII\ complex within 4434-4684\,\AA \\
EW\_FE\_4434\_4684\_ERR & DOUBLE  & \AA  & Measurement error in EW\_FE\_4434\_4684 \\
LOGBH\_HB\_VP06  & FLOAT   &  $[M_\odot]$  & Single-epoch BH mass estimate based on broad \hbeta\ \citep{Vestergaard_Peterson_2006} \\
LOGBH\_HB\_VP06\_ERR & FLOAT &  $[M_\odot]$  & Measurement error in LOGBH\_HB\_VP06 (systematic error not included) \\
\hline
\hline\\
\end{tabular}
}
\begin{tablenotes}
      \small
      \item NOTE. --- Format of the fits table containing the \hbeta-\OIII\ region spectral measurements of our sample. The full table is available in FITS format in the online version of the paper. 
\end{tablenotes}
\end{table*}

\subsection{Composite Spectrum of the Near-IR Sample}\label{sec:composite}

We create a composite spectrum of our near-IR sample combining optical SDSS and near-IR spectroscopy. We follow \citet{VandenBerk_etal_2001} to create a median composite spectrum, which better preserves the relative strengths of emission lines. The full composite spectrum is shown in Fig.\ \ref{fig:fullspec} and compared to the SDSS quasar composite in \citet{VandenBerk_etal_2001}. Our composite is slightly bluer than the \citet{VandenBerk_etal_2001} composite, which may be caused by our selection of objects with relatively blue continua typical of classical quasars, but a more likely explanation is that our luminous quasars are much less affected by host contamination than the low-$z$ and low-luminosity quasars used in the Vanden Berk et al.\ composite at rest-frame optical wavelengths \citep[see discussions in][]{Shen_etal_2011}. One advantage of our composite is that all objects contributed to the wavelength coverage, whereas the Vanden Berk et al.\ composite used high-$z$/high-luminosity quasars to cover the rest-frame UV and low-$z$/low-luminosity quasars to cover the rest-frame optical. This explains why the two composite spectra have similar UV broad-line properties but different optical line properties. The use of the Vanden Berk et al.\ composite for high-$z$ quasars should caution on the potential impact of host contamination in the composite.

Although we only have a small number of objects contributing to the rest-frame optical regime, we clearly detect several weak narrow lines such as \OI\ and \SII, which will be used to probe the physical properties of the NLRs of these high-$z$ quasars in future work. These detected weak narrow lines appear to be substantially broader than those in the composite spectrum generated for low-$z$ quasars. In addition, the broad \hbeta\ and \halpha\ lines also appear broader than the low-$z$ composite. These results are consistent with the fact that our near-IR sample represents the most luminous quasars and thus likely more massive hosts than low-$z$ quasars, hence both the broad lines and the narrow lines have larger widths than their low-$z$ counterparts with less massive BHs and hosts. 

The full composite spectrum for our near-IR sample is tabulated in Table \ref{table:composite}.

\subsection{Strength and Kinematics of \OIII\ }

We were able to measure \OIIIb\ for all but one of our objects, although for $\sim 35\%$ of them \OIIIb\ is not detected at $>3\sigma$ significance. Only for one object (J$0412-0612$) \OIIIb\ is not covered in our near-IR spectroscopy.  Fig.\ \ref{fig:oiii} (left) shows the \OIIIb\ REW as a function of continuum luminosity, where we also plot the low-$z$ SDSS quasars for comparison. The scatter in the \OIIIb\ REW is large for our high-$z$ quasar sample, but there is no obvious indication that the distribution is bimodal. The average \OIIIb\ strength is consistent with earlier near-IR spectroscopic studies on smaller samples of quasars with similar luminosities and redshifts as studied here \citep[e.g.,][]{Sulentic_etal_2004,Netzer_etal_2004}.

The average trend of decreasing \OIIIb\ REW with luminosity as shown in Fig.\ \ref{fig:oiii} (left) is known as the \OIII\ Baldwin effect \citep[e.g.,][]{Baldwin_1977,Brotherton_1996,Zhang_etal_2013,Stern_Laor_2012a,Stern_Laor_2012b}. \citet{Shen_Ho_2014} showed that the \OIIIb\ Baldwin effect is primarily driven by the flux reduction in the ``core'' component of \OIIIb\ when quasar luminosity increases. A byproduct of this effect is the increase in the overall \OIIIb\ width, as the broad ``wing'' component becomes more prominent at higher luminosities \citep[e.g.,][]{Shen_Ho_2014}. 

Recent near-IR spectroscopy of $z>1.5$ luminous quasars (unobscured or obscured) often report exceptionally large \OIII\ widths, with FWHM$\gtrsim 1000\,{\rm km\,s^{-1}}$ \citep[e.g.,][]{Netzer_etal_2004,Kim_etal_2013,Brusa_etal_2015}. Fig.\ \ref{fig:oiii} (right) shows the FWHM of the full \OIIIb\ profile as a function of continuum luminosity and compares with the low-$z$ SDSS quasar sample. The median \OIIIb\ FWHM is $\sim 1000\,{\rm km\,s^{-1}}$ for our near-IR sample, confirming the large FWHM values found in earlier work \citep[e.g.,][]{Netzer_etal_2004,Kim_etal_2013,Brusa_etal_2015}. However, despite the large scatter in our sample, they tend to follow the luminosity trend extrapolated from less luminous objects at lower redshifts. This suggests that the NLR kinematics of $z>1.5$ quasars is not significantly different from those at lower redshifts with similar quasar luminosities. 

To strengthen the above point, we use the median composite spectrum generated for our near-IR sample (\S\ref{sec:composite}), and compare it to the median composite spectra of low-$z$ SDSS quasars in different luminosity bins in Fig.\ \ref{fig:coadd}. When luminosity is matched, our high-$z$ sample shows a similar average \OIII\ profile as the most luminous low-$z$ quasars, suggesting there is limited redshift evolution in terms of \OIII\ properties when luminosity is matched. For comparison, in the bottom panel of Fig.\ \ref{fig:coadd}, we show the resulting median composite spectrum using the Hewett \& Wild redshifts for our high-$z$ quasars. The composite shows an additional broadening due to the systematic uncertainty in the Hewett \& Wild redshifts based on broad UV lines. In addition, the peak of \OIII\ in the Hewett \& Wild composite is blueshifted from that in the composite based on systemic redshifts by $\sim 2$\,\AA\ (rest-frame), consistent with the result in \S\ref{sec:spec_mea} and Fig.\ \ref{fig:zdiff} that the Hewett \& Wild redshifts (mostly \MgII-based) on average overestimate the \OIII-based systemic redshifts by $\sim 100\,{\rm km\,s^{-1}}$ for our near-IR sample. 

High-luminosity quasars display a strong blueshifted component in their \OIII\ emission \citep[e.g.,][]{Shen_Ho_2014}. We have used a simple decomposition method to decompose the \OIII\ emission into a core component and a blueshifted wing component. The median fraction of the blueshifted wing component to the total \OIII\ flux is $\sim 40\%$. The wing component for our near-IR sample shows a broad range of blueshift velocities of up to $\sim 1200\,\kms$, and has a median blueshift of $\sim 700\,{\rm km\,s^{-1}}$, much larger than the $\sim 200\,{\rm km\,s^{-1}}$ average blueshift of the wing component in much less luminous low-$z$ quasars ($L_{5100}\sim 10^{44}\,{\rm erg\,s^{-1}}$) \citep[e.g.,][]{Zhang_etal_2011,Shen_Ho_2014}. However, \citet{Shen_Ho_2014} showed that the blueshift velocity of the wing component increases with quasar luminosity, and the observed $\sim 700\,\kms$ velocity at the high luminosity regime sampled by our quasars ($L_{5100}\sim 10^{46}\,{\rm erg\,s^{-1}}$) is consistent with simple extrapolation from the luminosity trend found by \citet[][their Fig.\ E2]{Shen_Ho_2014}.

\subsection{Aperture Effects and NLR Sizes}

\begin{figure}
\centering
    \includegraphics[width=0.45\textwidth,height=0.1\textwidth]{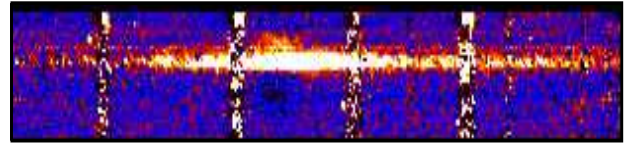}
    \caption{The rectified 2D spectrum in the \OIIIb\ region for J1220$+$0004 from FIRE. Vertical is the slit direction, which covers $\sim 6\arcsec$ along the slit. Horizontal is the wavelength direction, covering rest-frame 4970--5050\,\AA. The local continuum has been subtracted using a crude polynomial fit to pixels outside the \OIIIb\ region. Most of the \OIII\ flux is concentrated within the central $\sim 1\arcsec$. For this particular object there is evidence (the fluffy feature above the central emission) that some extended \OIII\ flux exists beyond the central $\sim 1\arcsec$, which, however, does not contribute to the total \OIII\ flux significantly. }
    \label{fig:2d_oiii}
\end{figure}

Our near-IR spectra were obtained using two different instruments: TripleSpec \citep{Wilson_etal_2004} on the ARC 3.5\,m telescope (38 quasars), and FIRE on the 6.5\,m Magellan-Baade telescope (36 quasars including the 14 quasars at $z\sim 3.3$). The TripleSpec observations used a slit width of either 1.1\arcsec\ or 1.5\arcsec, and the FIRE observations used a slit width of 0.6\arcsec. In both cases the slit was positioned at the parallactic angle in the middle of the observation. Since the major axis of the NLR is at random with respect to the slit position, our near-IR spectra on average enclose a physical region with comparable sizes to the slit width. Given the angular scale of $\sim 8.5\,{\rm kpc}/$\arcsec\ (at $z\sim 1.6$) and $\sim 7.5\,{\rm kpc}/$\arcsec\ (at $z\sim 3.3$), our near-IR spectra enclose all \OIII\ flux within $\sim 5-10\,{\rm kpc}$. Therefore the blueshifted wing \OIII\ component revealed in our spectra, if originated from quasar-driven outflows, is constrained to be within a few kpc from the nucleus. Future adaptive optics (AO) assisted near-IR IFU observations will improve the constraints on the spatial extent of this blueshifted \OIII\ component in luminous $z>1.5$ quasars.

We also expect that aperture losses due to the finite slit widths used are not important for our near-IR quasars. \citet{Netzer_etal_2004} showed that there is little \OIII\ flux beyond $\sim 1$\arcsec\ radius ($\sim 7\,{\rm kpc}$) in their slit near-IR spectra of high-$z$ quasars, which have similar luminosities and redshifts as our objects. With their careful analysis on the long-slit optical spectroscopy of $0.4<z<0.7$ type 2 quasars, \citet{Hainline_etal_2014} showed that there is an upper limit of $\sim 7\,{\rm kpc}$ on the size of the NLR in luminous quasars. Since the surface brightness of \OIII\ emission typically decreases with distance \citep[e.g.,][]{Liu_etal_2013a}, it is reasonable to expect that our aperture already encloses most of the \OIII\ flux, and the observed Baldwin effect is intrinsic and not subject to increasing aperture losses as quasar luminosity increases. 

To further ensure that we are not missing significant \OIII\ flux beyond our slit aperture, we visually inspected all 2D spectra for our objects. We found that in essentially all but 2--3 cases the \OIII\ flux is concentrated within the central $\sim 1\arcsec$ in the slit direction, and is consistent with being unresolved under the seeing conditions (e.g., similar spatial profiles for \OIII\ and continuum emission). In these 2--3 exceptions we see evidence that some extended \OIII\ emission exist beyond the central $\sim 1\arcsec$, but even in such cases the contribution of these extended emission to the total \OIII\ flux is negligible. An example of extended \OIII\ emission in our high-$z$ quasars is shown in Fig.\ \ref{fig:2d_oiii}. We therefore  confirm the earlier result in \citet{Netzer_etal_2004} that most of the \OIII\ flux in luminous $z>1.5$ quasars is within $\sim 10\,{\rm kpc}$ from the nucleus. However, to perform a detailed analysis on the spatial extent of the \OIII\ emission using our slit spectroscopy requires more careful spectral reductions/calibrations and a proper treatment of seeing effects, and is beyond the scope of the current work. 


\subsection{Eigenvector 1 Relations}

As mentioned in \S\ref{sec:intro}, another important spectral quantity that regulates  the strength and width of \OIIIb\ is the optical \FeII\ strength (e.g., the EV1 correlations). These spectral correlations are governed by simple underlying physical processes of the BH accretion. Early near-IR spectroscopic studies on small samples already hinted that similar EV1 correlations may exist in high-redshift quasars \citep[e.g.,][]{Sulentic_etal_2004, Sulentic_etal_2006,Runnoe_etal_2013b}.

\begin{figure}
\centering
    \includegraphics[width=0.49\textwidth]{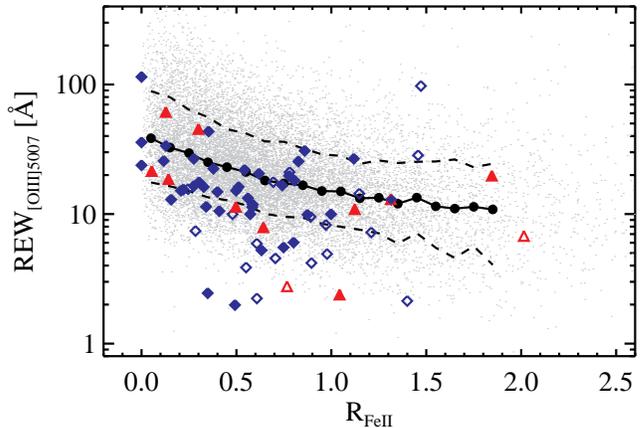}
    \caption{The anti-correlation between the \OIII\ EW and the optical \FeII\ strength (the EV1 relation). Symbol notations are the same as in Fig.\ \ref{fig:oiii}. The gray points are for the low-$z$ SDSS quasars in the \citet{Shen_etal_2011} catalog, and the black lines indicate the median, 16th and 84th percentiles of the distribution. Our $z>1.5$ quasars spread a similar range in optical \FeII\ strength, and follow the same EV1 trend defined by low-$z$ quasars. }
    \label{fig:rfe_oiii_ew}
\end{figure}

Fig.\ \ref{fig:rfe_oiii_ew} shows the anti-correlation between \OIII\ strength and optical \FeII\ strength, defined as $R_{\rm FeII}\equiv {\rm REW_{FeII}}/{\rm REW_{H\beta,broad}}$. Our near-IR quasars, although focused on the most luminous quasars, still shows a broad range of \FeII\ strength, and they fall consistently on the relation defined by the low-$z$ quasars (black lines), albeit with slightly lower \OIII\ REW given the Baldwin effect discussed earlier. Therefore we confirm earlier results \citep[e.g.,][]{Sulentic_etal_2004, Sulentic_etal_2006} that similar EV1 correlations already exist at $z>1.5$ with our much larger sample. 

\begin{figure}
\centering
    \includegraphics[width=0.49\textwidth]{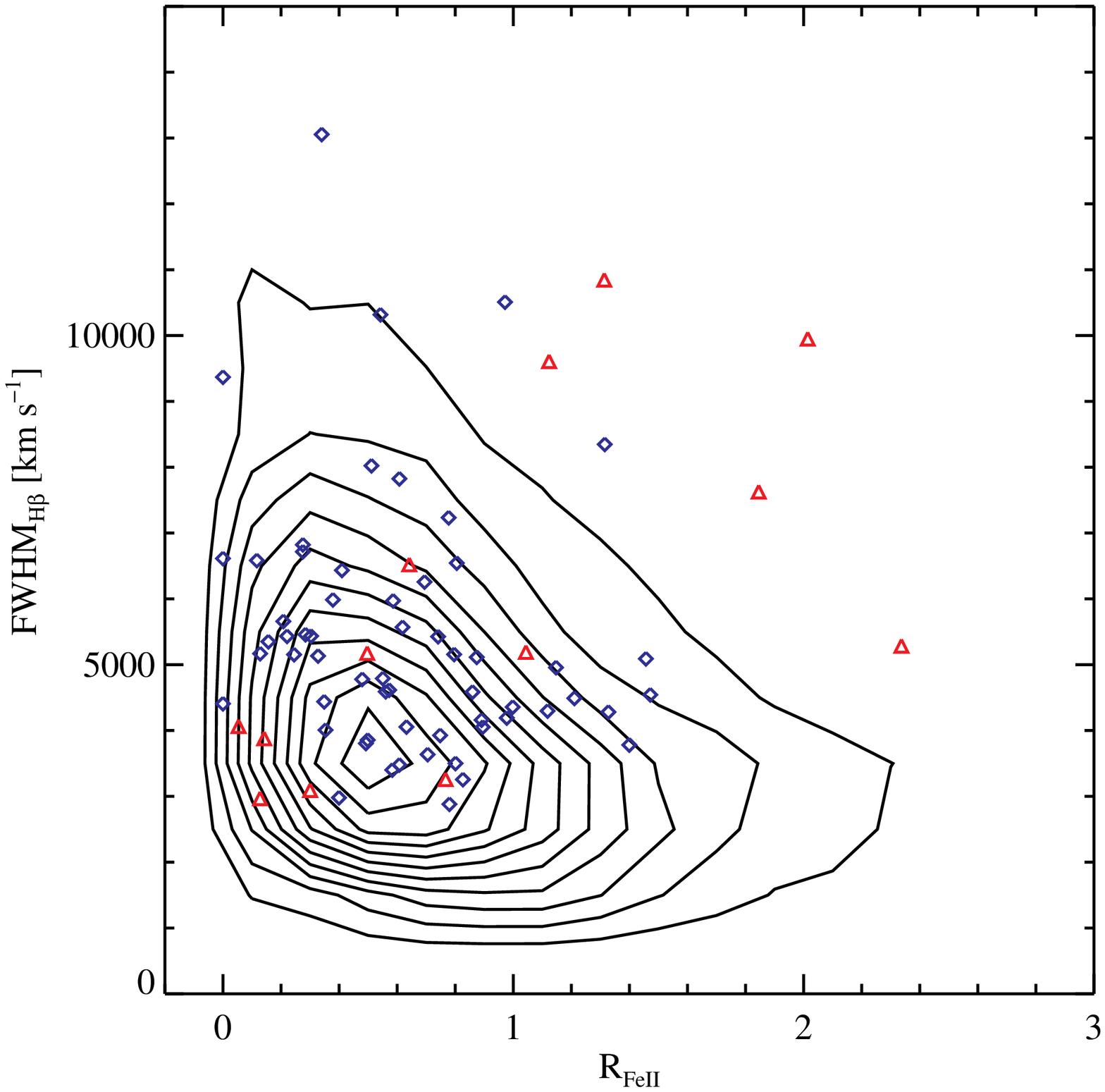}
    \caption{Distribution of our near-IR sample in the EV1 plane defined by $\Rfe$ and the broad \hbeta\ FWHM. The contours are for the low-$z$ and low-luminosity SDSS quasars based on the measurements in \citet{Shen_etal_2011}. The high-$z$ near-IR sample shows a similar wedged distribution, but the broad \hbeta\ FWHMs are offset to systematically larger values, which is consistent with the scenario that these most luminous quasars also have more massive BHs than their low-luminosity counterparts. Symbol notations are the same as in Fig.\ \ref{fig:dist}.}
    \label{fig:ev1}
\end{figure}

Fig.\ \ref{fig:ev1} shows the locations of our near-IR quasars in the 2D EV1 plane defined by the \FeII\ strength $R_{\rm FeII}$ and the broad \hbeta\ FWHM \citep[e.g.,][]{Boroson_Green_1992,Sulentic_etal_2000a,Shen_Ho_2014}. Notably our high-luminosity near-IR quasars show a systematic offset in the broad \hbeta\ FWHM, compared to the low-$z$ SDSS quasars (contours). The median broad \hbeta\ FWHM is $\sim 5000\,{\rm km\,s^{-1}}$ for our high-$z$ sample and $\sim 4000\,{\rm km\,s^{-1}}$ for low-$z$ SDSS quasars. This is consistent with our findings with the composite spectrum, and again indicates that these most luminosity quasars have larger BH masses than their low-$z$ and low-luminosity counterparts. We note that the vertical dispersion of objects in Fig.\ \ref{fig:ev1} is primarily an orientation effect, based on the results of low-$z$ quasars \citep[e.g.,][also see Marziani \etal\ 2001]{Shen_Ho_2014,Sun_Shen_2015}. Such an orientation bias in the broad \hbeta\ FWHM is also argued based on radio-loud quasars, where the orientation of the radio jet (hence the accretion disk and BLR orientation) can be inferred from the radio core dominance \citep[e.g.,][]{Wills_Browne_1986,Runnoe_etal_2013,Brotherton_etal_2015b}.

\section{Discussion}\label{sec:disc}

The strong blueshifted \OIII\ components and the exceptionally broad \OIII\ FWHMs observed in our broad-line quasars are consistent with recent studies of different types of AGN (e.g., ultraluminous infrared galaxies ULIRGs/AGN and obscured quasars, radio-loud and radio-quiet quasars) at similar redshifts and luminosities \citep[e.g.,][]{Harrison_etal_2012, Kim_etal_2013, Brusa_etal_2015}. In particular, \citet{Harrison_etal_2012} observed eight $z=1.4-3.4$ ULIRGs hosting AGN activity with near-IR integral field spectroscopy, and found evidence of broad \OIII\ emission on several kpc scales with velocity offsets of up to $\sim 850\,{\rm km\,s^{-1}}$. The FWHM, velocity offset and spatial extent of the \OIII\ emitting gas in their ULIRG/AGN sample are consistent with those inferred from our slit spectroscopy for luminous broad-line quasars (see \S\ref{sec:results}). Our results, along with these recent studies, suggest that kpc-scale outflows in ionized gas are common among the most luminous high-redshift actively accreting SMBHs. 

\citet{Brusa_etal_2015} presented near-IR spectroscopy for 8 obscured quasars at $z\sim 1.5$ and measured \OIII\ properties for 6 of them. Their obscured quasars have bolometric luminosities $\sim 10^{45-46.5}\, {\rm erg\,s^{-1}}$, about a factor of $\sim 10$ lower than those for our quasars, but are still among the luminous quasar population. Consistent with our results here, they reported large \OIII\ widths and velocity offsets for their small sample. They also performed a comprehensive comparison of the \OIII\ FWHM among different populations of active SMBHs, and found that their $z\sim 1.5$ obscured quasars have substantially larger \OIII\ FWHMs than those in $z<0.6$ type 2 quasars with similar \OIII\ luminosities. However, one potential caveat is that the \OIII\ properties in quasars (strength and profile) are also strong functions of continuum luminosity and EV1 \citep[e.g.,][]{Shen_Ho_2014}, so systems with matched \OIII\ luminosities may still have different physical properties such as quasar continuum luminosity or Eddington ratio.\footnote{Different selections, e.g., \OIII-based low-$z$ type 2 quasar selection versus X-ray selected high-$z$ obscured quasars, may also introduce additional complications in the comparison of their sample \OIII\ properties. } When matching the quasar continuum luminosity for our high-$z$ quasars and for low-$z$ SDSS quasars, we do not observe difference in their \OIII\ properties, suggesting negligible redshift evolution in the \OIII\ properties, at least in the most luminous unobscured broad-line quasars. 

We note that while significant \OIII\ flux beyond $\sim 10$ kpc is rare for our objects, there are exceptions at low redshift, such as extended \OIII\ emission at tens of kpc in both radio-loud and radio-quiet quasars (obscured and unobscured) \citep[e.g.,][]{Fu_Stockton_2009,Greene_etal_2011,Shen_etal_2011b,Fu_etal_2012,Husemann_etal_2013}, some of which may be due to mergers. Since our slit spectroscopy may miss such extended \OIII\ emission along other directions, a systematic search for extended \OIII\ emission in the general population of $z>1.5$ quasars with near-IR IFU observations is highly desirable.

It is interesting to note that a recent study of cool gas in quasar hosts at $z\sim 1$ traced by \MgII\ absorption imprinted on background quasar spectra  \citep{Johnson_etal_2015} also revealed a luminosity dependence of the \MgII-absorption gas covering fraction and velocity offset. Although the spatial extent of the \MgII-absorption gas in these low-$z$ quasars is much larger than that of the \OIII\ emission probed by our near-IR spectroscopy, and the luminosities of these $z\sim 1$ quasars are much lower than those of our high-$z$ quasars, it is possible that there is a connection between gas outflows on $\sim {\rm kpc}$ scales and on larger scales, as suggested by the similar luminosity trends seen in the two studies with different gas tracers. 

\section{Conclusions}\label{sec:con}

We have performed a detailed study on the rest-frame optical properties (focusing on the \hbeta-\OIII\ region) of $1.5<z<3.5$ luminous ($L_{\rm bol}=10^{46.2-48.2}\, {\rm erg\,s^{-1}}$) broad-line quasars, using a large sample of 74 objects with our own near-IR spectroscopy. The findings from this study are the following:

\begin{enumerate}

\item[$\bullet$] The redshifts of these high-$z$ quasars based on the UV broad lines (mostly \MgII) are uncertain by $\sim 200\,{\rm km\,s^{-1}}$ compared to the more reliable systemic redshifts from the peak of the narrow \OIII\ lines. In addition, the improved redshifts for SDSS-DR7 quasars by \citet{Hewett_Wild_2010} using broad UV lines are systematically biased high by $\sim 100\,{\rm km\,s^{-1}}$ from the \OIII-based redshifts for our quasars. 

\item[$\bullet$] The \OIII\ strength is lower than that for typical SDSS quasars at $z<1$, with a median REW of $\sim 13\,$\AA. Our high-$z$ objects tend to follow the same Baldwin effect of decreasing \OIII\ REW with quasar continuum luminosity as defined by low-$z$ quasars. 

\item[$\bullet$] The \OIII\ profile of these luminous quasars is highly asymmetric, with $\sim 40\%$ of the total flux in a blueshifted wing component on average. The wing component is on average blueshifted by $\sim 700\,{\rm km\,s^{-1}}$ from the systemic velocity. The overall \OIII\ width is exceptionally large, with a median ${\rm FWHM}\sim 1000\,{\rm km\,s^{-1}}$. These results confirm earlier observations with smaller near-IR spectroscopic samples at these redshifts \citep[e.g.,][]{Netzer_etal_2004}. 

\item[$\bullet$] However, we found that the strength and profile of \OIII\ of these high-$z$ luminous quasars are similar to those of their low-$z$ counterparts with comparable quasar continuum luminosity, and they follow the extrapolated trends with luminosity defined by the less luminous low-$z$ quasars. Therefore we conclude that the extreme properties of \OIII\ in these high-$z$ quasars are mainly driven by quasar luminosity rather than redshift evolution. 

\item[$\bullet$] Even within the limited dynamic range in quasar luminosity of our high-$z$ sample, we observe a similar spectral diversity in terms of the optical \FeII\ strength and the well known EV1 correlations for low-$z$ quasars. This suggests that the same physical processes that drive the diversity of quasars are already in place in these earlier active SMBHs. On the other hand, the average broad \hbeta\ FWHM is larger than that of the low-$z$ and lower-luminosity quasars, reflecting the larger BH masses in these high-$z$ quasars. 

\item[$\bullet$] Our slit spectroscopy suggests that most of the \OIII\ flux in our objects is within the central $\sim 10$ kpc, and the blueshifted \OIII\ wing component must also originate from below such spatial scales. We only found a handful of objects showing evidence of extended (but insignificant) \OIII\ emission beyond the central $\sim 10\,{\rm kpc}$ covered in our slit spectra, which will be good targets for spatially-resolved follow-up observations (such as adaptive optics assisted near-IR IFU observations).  

\end{enumerate}

The average values and spread in the \OIII\ REW and FWHM in luminous high-$z$ quasars presented here serve as a useful reference for planning near-IR spectroscopy to cover the \OIII\ region in high-$z$ quasars (e.g., to obtain a reliable redshift estimate based on \OIII). The relatively weaker (due to the Baldwin effect) and broader width of \OIII\ of these high-luminosity quasars compared to typical low-$z$ and low-luminosity quasars means that it is more difficult to detect and measure \OIII\ accurately for these luminous objects. 

We have concluded that these luminous $1.5<z<3.5$ quasars are not different from their low-$z$ counterparts at similar quasar continuum luminosities, in terms of the \OIII\ properties. The diversities in \OIII\ strength and kinematics are already clearly seen in recent studies using $z<1$ SDSS quasars \citep[e.g.,][]{Stern_Laor_2012a,Stern_Laor_2012b, Zhang_etal_2011, Zhang_etal_2013, Shen_Ho_2014}. In particular, we point out that blueshifted \OIII\ components are not unique to the most luminous quasars -- they are ubiquitous among quasars, with their properties (e.g., the fraction to total \OIII\ flux, velocity offset and width) correlated with quasar parameters \citep[luminosity and Eddington ratio, e.g.,][figs.\ E1 and E2]{Shen_Ho_2014}. Many recent studies use the kinematics of \OIII\ in different types of active galaxies to argue for AGN-driven outflows and feedback. Therefore it is important to understand the properties of \OIII\ emission in the general context of quasar parameter space, in order to understand the physical mechanisms driving these outflows. For example, the correlations of \OIII\ profile with both quasar continuum luminosity and optical \FeII\ strength \citep[see fig.\ 2 and fig.\ E2 in ][]{Shen_Ho_2014} suggest that simple accretion parameters (luminosity and Eddington ratio) may play the primary role in regulating the behaviors of \OIII\ outflows. 


\acknowledgements  


I thank the referee for comments that led to improvement of the manuscript, and Xin Liu and Luis Ho for useful discussions. Funding for the SDSS and SDSS-II has been provided by the Alfred P. Sloan
Foundation, the Participating Institutions, the National Science Foundation,
the U.S. Department of Energy, the National Aeronautics and Space
Administration, the Japanese Monbukagakusho, the Max Planck Society, and the
Higher Education Funding Council for England. The SDSS Web Site is
http://www.sdss.org/.

{\it Facilities}: Sloan, Magellan:Baade (FIRE), ARC (TripleSpec)


\end{document}